\documentclass[aps,prl,superscriptaddress,floats,showpacs]{revtex4}
\usepackage{amsmath,amssymb,ulem,graphicx}
\usepackage{color}

\begin{document}

\title{Probing the quantum state of a 1D Bose gas using off-resonant light
scattering}

\author{A. G. Sykes}
\affiliation{Jack Dodd Centre for Quantum Technology, Department of Physics, 
University of Otago, PO Box 56, Dunedin, New Zealand}
\affiliation{Theoretical Division and Center for Nonlinear Studies, Los Alamos
National Laboratory, 
Los Alamos, NM 87545, USA}

\author{R. J. Ballagh}
\affiliation{Jack Dodd Centre for Quantum Technology, Department of Physics, 
University of Otago, PO Box 56, Dunedin, New Zealand}

\date{\today}

\begin{abstract}
We present a theoretical treatment of coherent light scattering from an 
interacting 1D Bose gas at finite temperatures. We show how this can 
provide a nondestructive measurement of the atomic system states. 
The equilibrium states are determined by the temperature and interaction
strength, and are characterized by the spatial density-density 
correlation function.  We show how this correlation function is
encoded in the angular distribution of the fluctuations of the scattered light
intensity, thus providing a sensitive, quantitative probe of 
the density-density correlation function and therefore the quantum 
state of the gas. 
\end{abstract}

\pacs{pacs numbers}

\maketitle

The remarkable progress made in cold atom physics over the past two decades has 
been facilitated by the precision and flexibility of the measurement tools 
available to experimentalists. Optical probing by lasers has proved to be one 
of the most important tools, for example absorption and phase contrast
imaging~\cite{Ketterle_probe} have been  extensively used to measure atomic
density in bosonic and fermionic systems. Proposals for
employing coherent light scattering from cold atoms are long standing~\cite{javanainen_ruo}. 
More than a decade ago superradiant light scattering 
from a Bose condensate was demonstrated, and  stimulated light scattering was used
to implement Bragg scattering of matter waves~\cite{Ketterle_super_and_bragg}. 
More recently Mekhov {\it et
al}~\cite{ritsch} have proposed light scattering as a probe of cold atoms in 
an optical lattice.

In this letter we present a theoretical treatment of  monochromatic light
scattering appropriate for a 1D Bose gas. We show that for far detuned incident
light, a clear signature of  various quantum states of the gas can be obtained
from the fluctuations of the scattered light.  The 1D Bose gas  is an important
model in many-body physics, providing  one of the simplest paradigms of a
strongly interacting system, owing to its exact integrability via the Bethe
ansatz~\cite{liebliniger}. In addition, the distinction between 
fermions and bosons (that are well known in 3D systems), are suppressed
in 1D~\cite{cheon_shigehara}. 
For a  1D Bose gas with repulsive 
interactions, the system state can be characterized by
the two body spatial density-density correlation function~\cite{correlations1}.
Possible system states range from a nearly ideal gas in the `high' temperature
regime (where the spatial correlation function displays bunching),  to the
Tonks-Girardeau gas  in the interaction dominated regime, (where the correlation
function displays antibunching).  Experimental investigations of the 1D Bose gas
have been pursued in recent years~\cite{bec_low_dim_expts}. 
A method for measuring momentum 
correlations of a cold atom system has been demonstrated by 
analysing the density fluctuations of a resonant absorption image~\cite{Altman2004_GreinerJin2005}. 
Two and three body 
correlations were also studied via resonant absorption images~\cite{karen_expts}. 
Impressive as these experiments are, this technique cannot image submicron density fluctuations. 
In this work we show that a nondestructive 
measurement of the spatial nonlocal density-density correlation function of a  1D Bose
gas can be made by measuring the angular distribution of the fluctuations in the 
scattered light intensity.

We consider a system of two-state bosonic atoms in a long thin cylindrical trap,
illuminated by a coherent laser beam of frequency $\omega_{\rm L}$ and 
wavevector ${\bf k}_{\rm L}$. Each atom has resonant frequency $\omega_0$ and
transition dipole moment ${\mathbf d}$. We use the rotating wave
approximation for the radiation interaction. The laser mode is treated as a
classical quantity ${\mathbf E_{\rm L}}e^{i\left({\bf k_{\rm L}}\cdot
\mathbf{x}-\omega_{\rm L}\right)t}$, and its interaction  with the atoms is
characterised by the detuning $\Delta_{{\rm L}}=\omega_{{\rm L}}-\omega_0$ and
the Rabi frequency $\Omega_{{\rm L}}={\mathbf d}\cdot{\mathbf E_{\rm L}}/\hbar$.
In the case where the detuning is large, $\Delta_{{\rm L}}\gg\Omega_{\rm
L},\Gamma$, ($\Gamma$ is the spontaneous emission rate) the  upper
internal atomic state can be adiabatically eliminated, leaving coupled equations
for the scattered radiation modes $\hat{a}_{\mathbf{k}}(t)$ and the atomic 
field operator for the lower internal state  $\hat{\Psi}({\bf
x})$. The equation for $\hat{a}_{\mathbf{k}}(t)$ can be integrated, 
(ignoring small terms and employing the Markov approximation) leading to 
\begin{equation}\label{elim_light}
\hat{a}_{\mathbf{k}}(t)=
\frac{\pi \Omega_{\rm L} g_{\mathbf{k}}}{\Delta_{\rm L}}e^{-i\omega_{\rm
L}t}\delta(\Delta_{\mathbf{k}})e^{i\Delta_{\bf k}t/2}
\int_{\mathbb{S}^3}\! d^3\!\mathbf{x}\;
e^{i{\bf q}\cdot {\bf x}}\hat{\rho}({\bf x},t).
\end{equation}
Here  $g_{\mathbf{k}}$ denotes the usual  atom-radiation coupling constant,
$\Delta_{\mathbf{k}}= \omega_{\rm L}-\omega_{\mathbf{k}}$,
$\hat{\rho}=\hat{\Psi}^\dagger\hat{\Psi}$ is the 3D density of the system, ${\bf
q}={\bf k}_{\rm L}-{\bf k}$ and  $\mathbb{S}^3$ is the region of space where
the 
incoming light intersects the gas. We have excluded the additive quantum
noise term from Eq.~\eqref{elim_light} as we will only consider normally ordered
averages of $\hat{a}_{\mathbf{k}}$. 

For the cylindrically symmetric trapping potential
$V(\mathbf{x})=m\omega_\perp^2(y^2+z^2)/2$, the 1D regime can be achieved when
$\hbar\omega_\perp\ll k_{\rm B}T,\mu$ (the chemical potential).  In addition the atomic recoil energy
should  be much less than $\hbar\omega_\perp$. In
this case the atomic field operator is well approximated by
$\hat{\Psi}(\mathbf{x},t) =(\sqrt{\pi}
l_\perp)^{-1}\exp[{-(y^2+z^2)/2l_\perp^2}-i \omega_\perp
t]\hat{\phi}(x,t)$, where $\hat{\phi}(x,t)$ is the 1D annihilation operator of
the atomic ground state and $l_\perp=\sqrt{\hbar/(m\omega_\perp)}$ is the
transverse oscillator length. Performing the transverse integration in
Eq.~\eqref{elim_light} gives
\begin{equation}\label{light_1D}
 \hat{a}_{\mathbf{k}}(t)=A\;e^{-i\omega_{\rm L}t}\int_{\mathbb{S}^1}\!
dx\;\hat{n}(x)e^{iq_{\rm x}x}
\end{equation}
where $A={\pi \Omega_{\rm L} g_{\mathbf{k}}e^{i\Delta_{\bf
k}t/2}\delta(\Delta_{\mathbf{k}})e^{-(q_{\rm y}^2+q_{\rm z}^2)
l_{\perp}^2/4}}/{\Delta_{\rm L}}$, $\hat{n}=\hat{\phi}^\dagger\hat{\phi}$ is the
1D density and $\mathbb{S}^1=[-L/2,L/2]$ is the region where the light
intersects the gas. In this paper,  we consider the case where the   incident
light is a plane wave polarized along the $z$ axis, with ${\bf k_{\rm L}}$ in
the $x$-$y$ plane and  making an angle $\theta_{\rm L}$ to the $y$ axis. The
intensity of the scattered light is proportional to the total number of photons
in a given mode, 
\begin{equation}\label{photon_number}
\langle \hat{a}_{\mathbf{k}}^\dagger \hat{a}_{\mathbf{k}}^{\phantom{\dagger}}
\rangle=\left|A\right|^2\left[N\!+\!
2n^2\!\!\int_0^{L}dr\;g^{(2)}(r)\cos\left(q_{\rm x} r\right)\left(L-r\right)\right],
\end{equation}
where $N=nL$ is the total number of atoms illuminated. 
The first term in the square brackets in  Eq.~\eqref{photon_number} arises from
the atomic shot noise, while the integral provides the effect of the atomic
correlations, via the density-density correlation function of the gas, 
$g^{(2)}(x-x')=\langle:\!\hat{n}(x)\;\hat{n}(x')\!:\rangle/n^2$. We have assumed
that, within the illuminated region, the density is constant,
$\langle\hat{n}(x)\rangle=n$, and $g^{(2)}$ is translationally invariant.  

The  function $g^{(2)}$ characterises the state of a uniform density 1D
Bose gas~\cite{correlations1}.  At
equilibrium in the thermodynamic limit,  $g^{(2)}(r)$ is completely
determined by two parameters\cite{correlations2}: the dimensionless
interaction strength  $\gamma= m2\omega_\perp a/\hbar n$ (where $a$ is the 3D
s-wave scattering length, and $m$ is the atomic mass) and the reduced
temperature  $ \tau=T/T_{\rm d}$ where $T_{\rm d}=\hbar^2n^2/(2k_{\rm
B}m)$ is the temperature of quantum degeneracy. 
Varying these two parameters gives rise to three physically distinct
regimes,  each of which can be further split into two subregimes.

(i) {\it Nearly ideal-gas} regime, $\gamma\ll\min\{\tau^2,\sqrt{\tau}\}$. Here
the temperature always dominates over the 
interaction strength, and the local density-density correlation displays thermal
bunching, $1<g^{(2)}(0)\leq2$. This regime has two subregimes defined 
by $\tau\ll1$ (quantum decoherent gas) or $\tau\gg1$ (classical decoherent gas).

(ii) {\it Weakly interacting} regime,  $\tau^2\ll\gamma\ll1$. This regime, for
which both the interaction strength and the temperature are small, realises the
quasicondensate where density fluctuations are strongly suppressed but long
wavelength fluctuations of the phase still exist~\cite{quasiconds}. The
density-density correlation is close to the uncorrelated value of
$g^{(2)}(r)\approx1$, and fluctuations occur due to either vacuum or thermal
excitations. Two further subregimes can be defined, $\tau\ll\gamma$ (dominant
vacuum fluctuations) or $\tau\gg\gamma$ (dominant thermal fluctuations).

(iii) {\it Strongly interacting} regime (Tonks gas),
$\gamma\gg\max\{1,\sqrt{\tau}\}$. Here the interaction strength always dominates
over temperature induced effects, and the local density-density correlation
displays interaction-induced antibunching $0\leq g^{(2)}(0)<1$. This regime is
usefully divided into two further subregimes defined by $\tau\ll1$ (low
temperature Tonks gas) and $\tau\gg1$ (high temperature Tonks gas).

In each of the above regimes, perturbation theory can be applied  and six
different analytic expressions for the density-density correlation 
function have been calculated~\cite{correlations2}. The density--density
correlation function decays from its local value $g^{(2)}(0)$ to the
uncorrelated value $g^{(2)}=1$ over a length  $l_c$ ( the correlation length). 
For the quantum decoherent gas we have $l_c=2/(\tau n)$ (the phase coherence
length), while for  the classical decoherent gas, $l_c=\sqrt{4\pi/(\tau n^2)}$
(the thermal de Broglie wavelength). For the weakly interacting quasicondensate
regime, (with either vacuum or thermal fluctuations dominant) we have
$l_c=1/\sqrt{\gamma n^2}$ (the healing length). For the low temperature Tonks
gas we have $l_c=1/n$ and finally for the high temperature Tonks gas we have
$l_c=\sqrt{4\pi/(\tau n^2)}$ 

The physically relevant cases for light scattering are where $L \gg l_c >
\lambda$. In order to distinguish the incoherent from the coherent scattering it 
is convenient to define $\eta(r)\equiv g^{(2)}(r)-1$,
Eq.~\eqref{photon_number} then becomes
\begin{eqnarray}\label{expand_phot_nbr}
\langle \hat{a}_{\mathbf{k}}^\dagger \hat{a}_{\mathbf{k}}^{\phantom{\dagger}}
\rangle&=&\langle \hat{a}_{\mathbf{k}}^\dagger\rangle \langle
\hat{a}_{\mathbf{k}}^{\phantom{\dagger}}
\rangle + \left|A\right|^2 N \left[1+2\frac{N}{L^2}\int_0^{L}dr\;\eta(r)\cos\left(q_{\rm x}
r\right)\left(L-r\right) \right].
\end{eqnarray}
The
first term on the right of Eq.~\eqref{expand_phot_nbr}, $\langle
\hat{a}_{\mathbf{k}}^\dagger \rangle
\langle\hat{a}_{\mathbf{k}}^{\phantom{\dagger}}\rangle$, is the coherent
scattering and is
predominantly within angular width $\sim\lambda/L$ of the $q_{\rm x}=0$ direction. 
For $\lambda\ll L$,  the total intensity of coherently
scattered light at a distance $r$ from the gas is
\begin{equation}
I_c=N^2\left(\frac{\Omega_{\rm L}}{2\Delta_{\rm L}}\right)^2\Gamma\hbar\omega_{\rm L}\frac{3\sin^2\!\theta}{8\pi r^2}
{\rm sinc}^2\!\left[\frac{L\Theta}{2}\right],
\end{equation}
where $\Theta=k_{Lx}-k_{\rm L}\sin\theta\cos\theta_s$, and $\theta_{s}=\arctan(\frac{k_{x}}{k_{y}})$ is 
the {\it in-plane} scattering angle, and $\theta=
\arctan(\frac{\sqrt{k_{x}^2+k_{y}^2}}{k_z})$ is the 
{\it out-of-plane} scattering angle (see Fig.~\ref{fig_intensityisosurf}).

The signature for the atomic $g^{(2)}(r)$ function is in the {\it fluctuations} of the
scattered intensity arising from the integral term of
Eq.~\eqref{expand_phot_nbr}. These fluctuations are typically expressed in terms
of quadratures of the radiation field operator (relative to a particular phase
reference $\beta$), which are defined by
$\hat{X}_\beta^{{\mathbf{k}}}=\frac{1}{2}\left(\hat{a}_{\mathbf{k}}e^{-i\beta}
+\textrm{h.c.}\right)$. The variances of these quadrature operators,
\begin{eqnarray}\label{rfull}
&R_{\beta}^{\mathbf{k}}&=
\langle\left(\hat{X}_\beta^{{\mathbf{k}}}\right)^2\rangle-
\langle\hat{X}_\beta^{{\mathbf{k}} }\rangle^2 \nonumber  \\
&\phantom{R_{\beta}^{\mathbf{k}}}&=|A|^2N\left[\frac{1}{2}\!+\!\frac{N}{L^2}\int_0^Ldr\; \eta(r)
\left(\cos(q_{\rm x}r)(L\!-\!r)+\cos(2\beta)
\frac{\sin\left[q_{\rm x}(L-r)\right]}{q_{\rm x}}\right)\right],
\end{eqnarray}
can be measured by well known techniques. 
For $L\gg l_c >\lambda$, the dependence on 
$\beta$ in Eq.~\eqref{rfull} is very tightly concentrated around the 
transmitted and reflected directions; i.e. $|q_{\rm x}|\lesssim 1/L$. 
A simple and prudent choice of phase reference is $\beta=\frac{\pi}{4}$. 
In this case the quadrature noise simplifies to
$R_{\frac{\pi}{4}}^{\mathbf{k}}\approx|A|^2
N\left[1/2+n\int_0^L\!dr\;\eta(r)\cos(q_{\rm x}r)\right]$, which consists
of the atomic shot noise term, and a Fourier transform of $\eta(r)$. 

In order to obtain an analytic understanding of our results, we  first take the
limiting cases where $\gamma$ and $\tau$ take on 
values which lie {\it deep} within one of the regimes defined above. The
density-density correlation then simplifies to
\begin{equation}
\eta(r)\approx\left\{
\begin{array}{ll} 
\exp\left[-\tau n r\right] & \qquad\textrm{(a)}\\
\exp\left[-\tau n^2 r^2/2\right] & \qquad\textrm{(b)}\\
-\exp\left[-\tau n^2 r^2/2\right] & \qquad\textrm{(c)}\\
-\sin^2(\pi n r)/(\pi n r)^2 & \qquad\textrm{(d)}
\end{array}
\right.\nonumber
\end{equation}
(a) is deep within the decoherent quantum regime, 
(b) is deep within the decoherent classical regime, 
(c) is  deep within the high temperature Tonks regime, 
(d) is  deep within the low temperature Tonks regime. 
The integral for $R_{\frac{\pi}{4}}^{\mathbf{k}}$ can then be calculated analytically;
\begin{equation}
 R_{\frac{\pi}{4}}^{\mathbf{k}}=\frac{|A|^2N}{2}\left\{
\begin{array}{ll} 
1+2 n^2\tau /(q_{\rm x}^2+n^2\tau^2) & \qquad\textrm{(a)}\\
1+2\sqrt{\pi/(2\tau)}e^{-q_{\rm x}^2/(2n^2\tau)} & \qquad\textrm{(b)}\\
1-2\sqrt{\pi/(2\tau)}e^{-q_{\rm x}^2/(2n^2\tau)} & \qquad\textrm{(c)}\\
1-\textrm{tri}\left[q_{\rm x}/(2\pi n)\right]& \qquad\textrm{(d)}\\
\end{array}
\right.\label{eq:Rlimits}
\end{equation}
where $\textrm{tri}(r)=\max(1-|r|,0)$. We note that for $\eta =0$, the case of a
perfectly coherent quasicondensate, density fluctuations are completely
suppressed, and  $R_{\frac{\pi}{4}}^{\mathbf{k}}$ reduces to the shot noise. For
intermediate values of $\gamma$ and $\tau$, we solve for $R_\beta^{{\bf
k}}$ numerically using expressions for $g^{(2)}$ taken from~\cite{correlations2}. As 
seen in Fig.~\ref{fig_intensityisosurf}, the intensity
fluctuations arising from $\eta(r)$ are typically scattered over a wider angle
than the coherent scattering. For example, the thermal case in 
Eq.~\eqref{eq:Rlimits}(b), the angular spread is
$\sim\arctan\left(N(\lambda/L)(2.3\sqrt{\tau}/2\pi)\right)$. 
As a crude approximation, we can write $\eta(r)=\eta e^{-r^2/(2l_c^2)}$, where 
$-1\leq\eta\leq1$, and we find the incoherent scattering intensity, at a distance $r$ 
from the gas, to be $I_{ic}=$
\begin{equation}
N\left(\frac{\Omega_{\rm L}}{2\Delta_{\rm L}}\right)^2\Gamma\hbar\omega_{\rm L}\frac{3\sin^2\!\theta}{8\pi r^2}\left[1+
nl_c\eta\sqrt{2\pi}\exp\!\!\left(\frac{-l_c^2\Theta^2}{2}\right)\right].
\end{equation}
Intensity isosurfaces of both the coherent and incoherent scattering are shown in 
Fig.~\ref{fig_intensityisosurf}.

\begin{figure}
 \includegraphics[width=0.6\columnwidth]{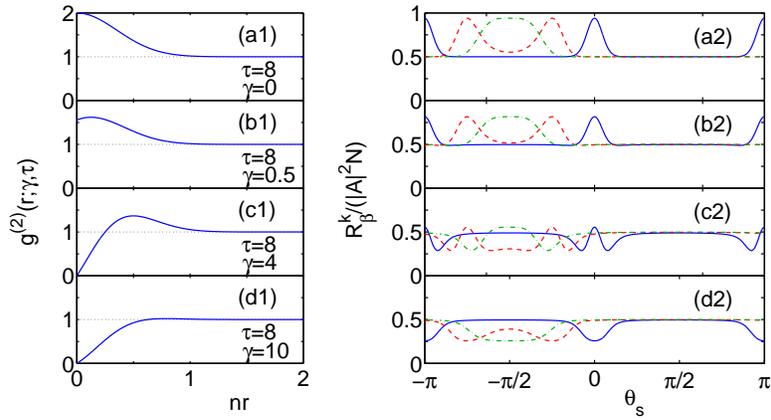}
\caption{(color online). Density-density correlation function (LH column) and  corresponding
scattered intensity quadrature noise 
 (RH column). Sequence shows transition from a decoherent classical gas to a 
high temperature Tonks gas. Incident angle $\theta_{\rm L} = 0$  (solid blue
lines); $\pi/4$ (red dashed 
lines); $\pi/2$ (green dot-dashed lines). Other parameters $({\bf k}_{\rm
L})_{\rm x}=20n$,  and $L\gg l_c$.}
\label{fig:DC_to_HighTTonks}
\end{figure}

\begin{figure}
 \includegraphics[width=0.6\columnwidth]{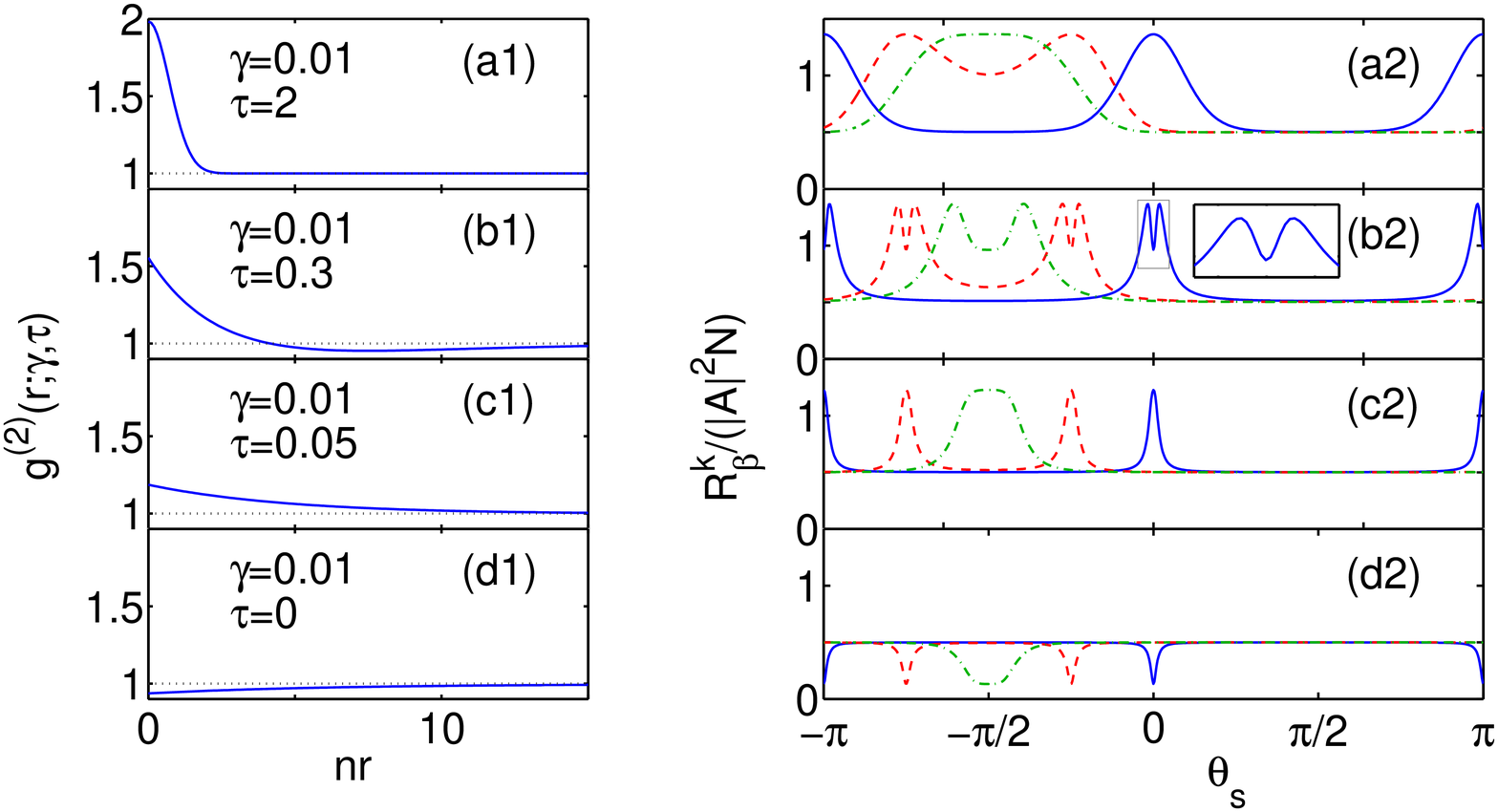}
\caption{(color online). Density-density correlation function and  scattered intensity
quadrature noise for the 
transition from a decoherent classical gas to a 
zero temperature quasicondensate. Incident angle $\theta_{\rm L} = 0$  (solid
blue lines); $=\pi/4$ (red dashed 
lines); $=\pi/2$ (green dot-dashed lines).   $({\bf k}_{\rm L})_{\rm x}=5n$, and 
$L\gg l_c$.}
\label{fig:DC_to_quasicond}
\end{figure}

\begin{figure}
 \includegraphics[width=0.6\columnwidth]{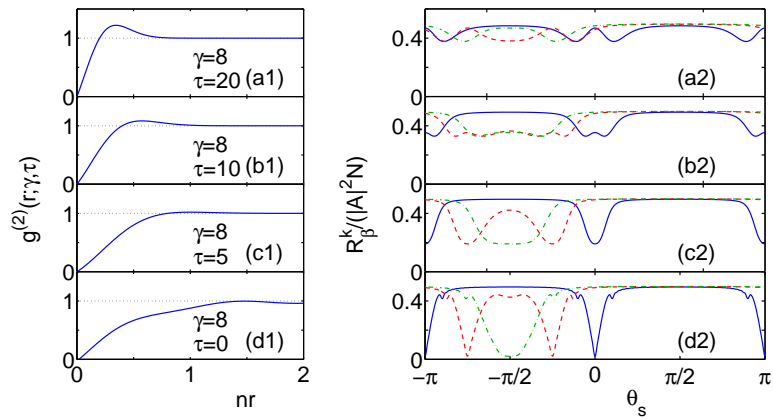}
\caption{(color online). Density-density correlation function and  scattered intensity
quadrature noise for the transition from a high temperature Tonks gas to a 
low temperature Tonks gas. Incident angle $\theta_{\rm L} = 0$  (solid blue
lines); $=\pi/4$ (red dashed 
lines); $=\pi/2$ (green dot-dashed lines).  $({\bf k}_{\rm L})_{\rm x}=20n$, 
$L\gg l_c$.}
\label{fig:HighTTonks_to_LowTTonks}
\end{figure}

\begin{figure}
\begin{center}
 \includegraphics[width=0.9\textwidth]{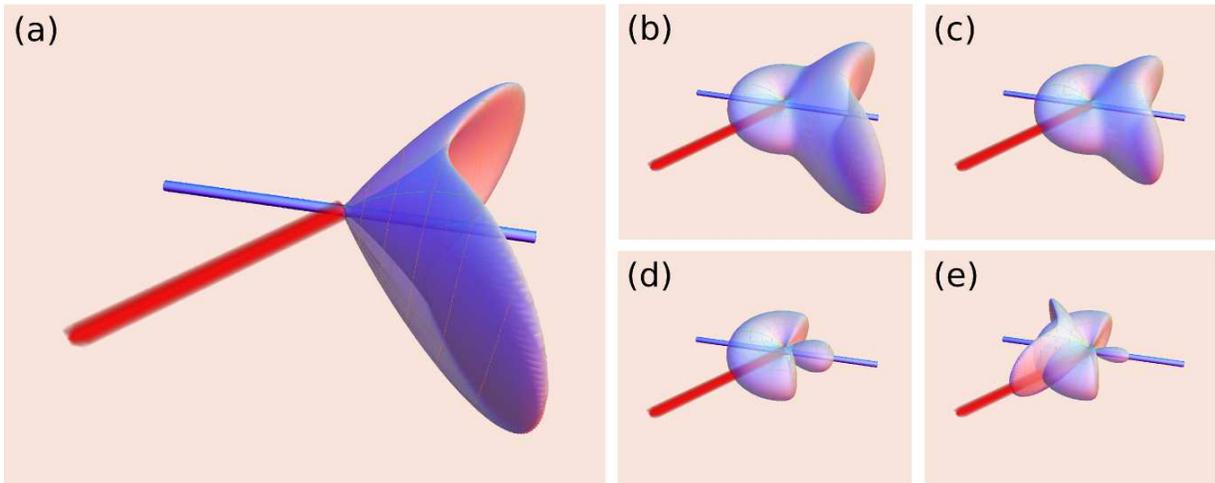}
\end{center}  
\caption{Intensity isosurfaces for coherently and incoherently scattered light. In each of the above figures, 
the blue cylinder shows the orientation of the 1D Bose gas, the red cylinder shows the 
direction of the incoming laser beam, and the semi-opaque, purple surface shows an intensity isosurface as a function of scattering angle. 
In (a) we show the intensity of the coherently scattered light [given by Eq. (5) in the main text] with 
$N^2\left(\frac{\Omega_{\rm L}}{2\Delta_{\rm L}}\right)^2\Gamma\hbar\omega_{\rm L}\frac{3}{8\pi r^2}=1$, $k_{\rm Lx}=0.7k_{\rm L}$ and 
$k_{\rm L}L\approx28$. In (b)--(e) we show the incoherent intensity [given by Eq. (8) in the main text], for the approximation 
$\eta(r)=\eta e^{-r^2/(2l_c^2)}$, with $N\left(\frac{\Omega_{\rm L}}{2\Delta_{\rm L}}\right)^2\Gamma\hbar\omega_{\rm L}\frac{3}{8\pi r^2}=1$, 
$k_{\rm Lx}=0.7k_{\rm L}$, $nl_c=1$ and $k_{\rm L}l_c=7$. In (b) we have $\eta=0.5$; in (c), $\eta=0.3$; in (d), $\eta=-0.3$; 
and in (e), $\eta=-1$.}\label{fig_intensityisosurf}
 \end{figure}

The results for $R_{\frac{\pi}{4}}^{{\bf k}}$ plotted against the scattered angle 
are shown in the right-hand columns of
Figs.~\ref{fig:DC_to_HighTTonks}--\ref{fig:HighTTonks_to_LowTTonks}, for a range
of $\tau$ and $\gamma$.  In the left hand-columns  we plot the corresponding
$g^{(2)}(r)$.  In Fig.~\ref{fig:DC_to_HighTTonks} the sequence of plots are at a
fixed high temperature, and show the transition from a decoherent classical gas
(top row) to a high temperature Tonks gas, as the interaction strength $\gamma$
increases from 0 to 10.
For weak interaction $g^{(2)}$ displays a peak at $r\approx0$; thermal
bunching. As $\gamma$ increases, the peak in $g^{(2)}$ changes
to a dip; antibunching. The behaviour of the corresponding
scattered intensity fluctuations, which consist of a noise floor ($|A|^{2}N/2$),
plus the Fourier transform of $\eta=g^{(2)}-1$, can be readily understood. A
positive (negative) function $\eta$ produces a peak (trough) in
$R_{\frac{\pi}{4}}^{{\bf k}}$, with a width scaling inversely to $l_c$.
This is clearly illustrated in the sequence of plots for $\theta_{\rm L}=0$ in
Fig.~\ref{fig:DC_to_HighTTonks}. Atom bunching produces enhanced intensity
fluctuations over a range of angles around $\theta_{s}=0$, while antibunching
results in noise suppression in the scattered light over a similar angular
range. We can interpret the more complex structure in Fig 1 (c2) as being due to
a remnant  bunching on a long spatial scale  superimposed on a more dominant
(and spatially narrower) antibunching. For an incident angle of $\pi/4$, these
features are
replicated around a (broader) transmitted peak at $\theta_{s}=-\pi/4$ and a 
reflected peak at $-3\pi/4$. 
Fig.~\ref{fig:DC_to_quasicond}, where all plots are at weak interaction strength
$\gamma = 0.01$, shows the transition from a decoherent classical gas 
to a zero temperature quasicondensate as temperature is decreased to zero. At
the higher temperatures, the scattered intensity fluctuations are enhanced in a
broad angular region around the transmitted (and reflected) beams. At the lowest
temperature, where $g^{(2)}$ displays weak antibunching with $l_s$ given by the
condensate healing length (which is long), the light fluctuations are suppressed
over a narrow angular range. Fig.~\ref{fig:DC_to_quasicond}(b1) exhibits
competition between bunching and antibunching on different length scales,
resulting in a subtle minima in $g^{(2)}$  at $nr\approx\tau/(4\gamma)$. This
leads to a sharp dip in the  noise features Fig.~\ref{fig:DC_to_quasicond}(b2) 
showing that  light scattering is an ideal probe of these subtle, and
controversial~\cite{correlations2}, density-density correlations. Finally, in
Fig.~\ref{fig:HighTTonks_to_LowTTonks}, a sequence of plots at constant large 
interaction strength $\gamma = 8$, shows the transition from a high temperature
to low temperature Tonks gas. In our analysis we have assumed the atoms are in
an equilibrium state unperturbed by the interaction with the light. This
assumption is valid provided the heating per atom from the recoil energy 
is much less than the initial energy per atom, $E_{0}$. The recoil
component perpendicular to the trap axis is absorbed by the trap as a whole
(because it is very tight), and so for $\theta_{\rm L}=0$, only the incoherently
scattered light contributes to the heating. We can estimate  the heating rate
per atom to be $\Upsilon\hbar^2 \langle k_{x}\rangle^{2}/2m$ where $\langle
k_{x}\rangle^{2}$ is the mean square recoil momentum  and
$\Upsilon=\Gamma(\frac{\Omega_{\rm L}}{2\Delta_{\rm L}})^{2}$  is the rate of photon scattering
\cite{CohenT}. The perturbation to the initial state of the gas will be
negligible provided $\Upsilon t\ll E_{0}$. 
Expressions for  $E_{0}$ can be found from~\cite{liebliniger}. 

In summary, we have shown that the quantum state of a 1D Bose gas can be 
nondestructively measured using off resonant light scattering. The angular 
distribution of the quadrature noise in the scattered light provides a direct
signature of the nonlocal two-body correlations which characterize different system
states of the gas. The method is useful across the full range of temperatures and
atomic interaction strengths, and is capable of revealing subtle features in the 
density fluctuations that occur across multiple length scales. In addition, the method 
could also be used to monitor the dynamics of two-body correlations 
in the system, by simply solving the inverse Fourier transform of the intensity 
fluctuations.

This work was supported by the New Zealand Foundation for Research, Science and
Technology under Contract No. NERF-UOOX0703. AGS gratefully acknowledges the 
support of the U.S. Department of Energy through the LANL/LDRD Program for this work.

\end{document}